\documentclass[aps]{revtex4}
\usepackage{amssymb}
\usepackage{graphicx}% Include figure files
\usepackage{bm}% bold math

\begin{document}
\title{The Unification of Four Fundamental Interactions}
\author{Hai-Jun Wang}
  \altaffiliation[To find all my favorite papers: ]{Please search
for author "Wanng" in the arXiv database.}

\address{Center for Theoretical Physics and School of Physics, Jilin
University, Changchun 130023, China}

\begin{abstract}
In this paper, we first incorporate the weak interaction into the
theory of General Nonlocality [J. Math. Phys. 49, 033513 (2008)]
by finding a appropriate metric for it. Accordingly, we suggest
the theoretical frame of General Nonlocality as the candidate
theory of grand unification. In this unifying scenario, the
essential role of photon field is stressed.
\end{abstract}
 \maketitle

\section{Introduction}
It has been decades since scientists commenced their efforts to
unify the four fundamental interactions. The efforts focused
mainly on how to quantize the gravity. In that context some new
theories such as String theory, Quantum Loop Gravity and
Non-Commutative Geometry have been developed. In contrast, as we
know, no attempt has been practiced in the opposite direction: to
unify the quantum theory under the frame of General Relativity.
Now such an opposite scheme is feasible. Since we find our
previous theory of General Nonlocality [1] can also include the
weak interaction as a part.

The remainder of the paper is arranged as follows. in section 2,
we first present the detailed analysis on where the fermion mass
should come from --the problem unresolved in our previous
manuscript[1], and subsequently the weak interaction is
incorporated into the theory of General Nonlocality in section 3.
Finally, we suggest the theoretical frame of General Nonlocality
as universal law of fundamental forces.

\section{The origin of mass of material particles}

In the restored Dirac equation (8.12) in Ref. [1], the required
mass term in Eq. (8.5) is missing. And we have tried to remedy
this by two methods there, but they are not satisfactory. One
method is to add a mass term to the right hand side of Eq. (8.9)
directly, then Eq. (8.2) would have a nonzero term on its right
hand side too. That obviously contradicts the hypothesis that
motion equation is just the geodetic line. The other method is to
accept the term $i\,\int A_\nu \partial _\lambda \partial ^\nu
{\psi \,}dx^\lambda $ as the mass term. However, on one hand this
term is path-dependent (and thus \textbf{nonlocal}) if the
integration is not over a closed loop; on the other hand, even if
the closed loop is performed, one notes that the coupling of
$A_\nu $ and ${\psi }$ would give the mass value no more than
$m/\sqrt{137}$ ($m$ is the electron mass). So this method is also
infessible. The origin of fermion mass is still a problem.

The mass problem arises after the projection from geodetic
equation (8.2) to its space-time representation (8.12) via the
replacement $d\rightarrow \gamma _\mu \partial ^\mu $. To respect
the hypothesis that motion equation is the geodetic line we may
not make any alteration to the starting point Eq. (8.2). Also, the
phrase ''plane wave'' appears in the introduction of section
VIII--A should not be confused with the conventional term in
quantum mechanics. In former case the ''plane wave'' means the
local plane wave $d{\psi \,}$, after the electron wave is observed
by another fermion \textbf{locally} in a \emph{particular
complex-frame} (such frame is assumed to always exist). So putting
a plane wave $e^{i\vec k\cdot \vec x-i\omega t}$ (in the common
sense of quantum mechanics) into Eq. (8.2) to check the mass term
is inappropriate.

Obviously, the replacement $d\rightarrow \gamma _\mu \partial ^\mu
$ is invariant under Lorentz transformation. However, if the same
Lorentz group is viewed as the structure group obeyed by local
plane wave $d{\psi }$, then the parallel displacement as well as
the motion equation (8.2) may not exist any longer, since the
existence of the \emph{particular frame} can not be guaranteed
under the mere transformations of complex representation of
Lorentz group $SL(2,\not C)$--$D^{(1/2,1/2)}$. Therefore, the
transformation of Eq. (8.2) has nothing to do with the mass term
either.

Summarizing the above analyses, the mass problem must lie in the
replacement $d\rightarrow \gamma _\mu \partial ^\mu $ and the form
of plane wave we substitute in Eq. (8.12). Therefore the mass
origin has to do with real space instead of complex space. In
complex space, the mass differences become trivial for material
particles. So, the origin of the fermion mass may lie in
the more delicate replacement than $d\rightarrow \gamma _\mu \partial ^\mu $%
. If we refine the replacement by

\begin{eqnarray*}
d^{+} &\rightarrow &\gamma _\mu \partial ^\mu -im \\
d &\rightarrow &\gamma _\mu \partial ^\mu +im \\
dd &\rightarrow &d^{+}d
\end{eqnarray*}
then we obtain the satisfactory dominating terms $(\Box +m^2)\psi
,$ but in the second term of Eq. (8.9) several other redundant
terms would appear. Even if these redundant terms are not the
troublesome, we still face the fact of applying consistently the
same replacement to the derivation of the boson field equations,
which undoubtedly will ruin the previous formalism of the obtained
field equation as well as the perfect understanding of the boson
mass origin.

So, the refinement of the replacement is also limited.

Now let's return to physics. It can be noticed that fermion masses
always accompany the appearance of charges, with the exceptions
such as the neutrino and neutron, but the neutrino has almost no
mass and neutron has a relatively larger magnetic moment. So
mostly, if a fermion of spin-$1/2$ has nonzero mass, it must be
nontrivially relevant to the electromagnetism. In formalism
concerning special relativity, this relevance is best expressed by
the relationship between $\not
\partial =\gamma _\mu \partial ^\mu $ and $m$. So, we pose the
hypothesis that the appearence of the term $\gamma _\mu
\partial ^\mu $ is always accompanied by the mass term, both before the wave
function $\psi $. The further understanding of this judgement
roots in the fact that in micro-world, the fermion mass is
detected almost completely via electromagnetic interaction, in
contrast to our knowledge that in our everyday life mostly we
evaluate masses with the aids of gravitation. Based on this
understanding of the operator $\gamma _\mu \partial ^\mu $, we add
a mass term $-m^2\psi $ at will to the right hand of the equation
(8.9) once we begin to project it to space-time, as a part of
projection.

By ruling out several possibilities of adding mass term, finally
we are subject to a physical manner. So far, all the terms in
quadratic form of Dirac equation [1] can be restored from our
geodetic equation.

The understanding of origination of boson mass in the paper [1] is
consistent and perfect.

\section{The metric for the weak interaction}

Naturally, the method of nonlocality [1] is expected to describe
the weak interaction too. Here we use the known rules--the
definition of two sides of the boundary of physical region--to
carry out what the metric matrix should be for the weak
interaction. In the attempt we use the criteria (11.6)
\begin{equation}
\mid A_{\bar \alpha \beta }\mid =1-A_0^2+\vec A^2=\{
\begin{array}{cc}
0 & \text{bound states} \\
1 & \text{ asymptotically--free states}
\end{array}
\end{equation}
and approximation form (11.14a)

\begin{equation}
(A_{\bar \alpha \beta }^{ab})=\Gamma ^0\otimes I_{2\times 2}+A_\mu
\Gamma ^1\gamma ^\mu \otimes \vec A_a\tau ^a\text{ , }\tau
^a\text{ are the Pauli matrices}
\end{equation}
in Ref. [1] as the starting point (In this paper these two
equations denoted as Eq. (1) and Eq. (2) respectively.). The
$\Gamma ^0$ in the above equation is the \textbf{initial metric
matrix} we are searching for. To make the metric matrix satisfy
the
boundary condition Eq. (1), we have to search for the appropriate form of $%
\Gamma ^0$ in order to include matrix factor $\gamma ^\mu \gamma
^5$ in the $\Gamma ^1$ of Eq. (2). In what follows we list the
possibilities and give the discussions.

1. First we examine the possibility that in the absence of
interaction, if the (\textbf{initial}) metric matrix form is
$\gamma _0\,\gamma _5$, then after a period of interaction the
matrix evolves into $\gamma _0\gamma _5+\gamma _0\gamma _\mu
\gamma _5A^\mu $, i.e. the interaction vertex $\bar \psi \gamma
_5\psi \rightarrow \bar \psi \gamma _\mu \gamma _5\psi A^\mu $
(apart from a coupling constant). The explicit form of the metric
matrix is

\begin{equation}
(\gamma _0+\gamma _0\gamma _\mu A^\mu )\gamma _5=\left(
\begin{array}{cc}
-\vec \sigma \cdot \vec A & 1+A_0 \\
-1+A_0 & -\vec \sigma \cdot \vec A
\end{array}
\right) \text{ ,}
\end{equation}
which leads to the determinant
\begin{equation}
\det (\gamma _0\gamma _5+\gamma _0\gamma _\mu \gamma _5A^\mu )=\vec A%
^2+1-A_0^2\text{ ,}
\end{equation}
which is just the form of electrodynamics, meeting the
requirements of both strong interaction side (bound state) and
weak interaction side (asymptotically-free state). The choice of
$\gamma _0\,\gamma _5$ is a promising candidate for $\Gamma ^0$.
Whereas the actual vertex of
weak interaction has the form like $\bar \psi \gamma _\mu (1\pm \gamma _5)\psi A^\mu $%
, we should combine this vertex with \textbf{initial metric}
$\gamma _0\,\gamma _5 $.

2. Examine the possibility $\bar \psi \gamma _5\psi \rightarrow
\bar \psi \gamma _\mu (1\pm \gamma _5)\psi A^\mu $: the metric
matrix is
\begin{equation}
\gamma _0\gamma _5+\gamma _0\gamma _\mu A^\mu (1\pm \gamma
_5)=\left(
\begin{array}{cc}
X & 1\pm X \\
-1\pm X & X
\end{array}
\right) \text{ ,}
\end{equation}
where $X=A_0\pm \vec \sigma \cdot \vec A$. The corresponding
determinant yields
\begin{equation}
\det (
\begin{array}{cc}
X & 1\pm X \\
-1\pm X & X
\end{array}
)=1\text{ ,}
\end{equation}
which meets the requirement of asymptotically-free side only. As
well known, the weak interaction gives nonsupport of bound states.
So the Eq. (5) and the \textbf{initial metric} form $\gamma
_0\,\gamma _5$ are just the perfect ones.

The Eq. (6) automatically
holds regardless of the details of operator $X$%
, so we cannot infer any further conclusions from it, even if
substituting the concrete form of $\vec A$
\begin{equation}
\vec A=\vec A_a\tau ^a\text{ .}
\end{equation}

The other determinants such as $\det (\gamma _0+\gamma _0\gamma
_\mu (1\pm \gamma _5)A^\mu )$ and $\det (\gamma _0(1\pm \gamma
_5)+\gamma _0\gamma _\mu (1\pm \gamma _5)A^\mu )$ are also
examined, but they present no required boundary properties.

Additionally, extension of the group from $SU(2)$ to $U(2)$ is
still necessary while considering the curving effect of weak
interaction.

\section{Universal law for fundamental forces}

So far, we have included all of the three microscopic interactions
in the theoretical frame of General Nonlocality. Additionally, in
view of the well-known puzzles of conservation law of
energy-momentum tensor in General Relativity--which is
understandable if the energy-momentum tensor is also nonlocal, we
can put forward the hypothesis that \emph{the dynamics for all of
the fundamental forces should be described with the
\textbf{common} motion equation} (geodetic line) \emph{and field
equation} ($R=0$) \emph{in geometrical manner}, which
automatically induces the Nonlocality.

The history of the development of description of dynamics, after
Newton equation, have underwent several stages, as follows

\[
\text{Hamitonian}\longrightarrow \text{Lagarangian}\longrightarrow \text{%
Action}\longrightarrow \text{Geometry},
\]
each of the succeeding one is more general than its preceding one.
The Lagrangian is able to include the covariant form of special
relativity, and the action form can incorporate the gauge fixing
condition naturally and make the quantization process more fluent.
As for geometry, we hope it can circumvent the renormalization
processes, as well as provide other nonperturbative methods.

The curvings (Geometries) corresponding to the four fundamental
interactions are all relevant to effects of photon. It is well
known that the curvature of space-time (in General Relativity) has
been confirmed by the curving effect of light. But for the
microscopic interactions, one may not be aware of how does the
photon field curve in complex space--unitary space. We assert here
that the unifying picture of electrodynamics, weak interaction
(the weak doublets), and quark dynamics (color triplet) may be
achieved by regarding the curving of photon's field, just as
implied in Ref. [1]:

The photon exists as a pure energy form, which can decay into
lepton-antilepton pair$\rightarrow $proton-antiproton
pair$\rightarrow $quark-antiquark pair. The processes successively
take place with the increase of the photon energy. The remarkable
feature of the successive processes is that the lepton-antilepton
is in the $U(1,\not C)$ region of the gauge fields, and
proton-antiproton is in $U(2,\not C)$ hadron region [forms the
weak doublet together with neutron], and quark antiquark in
$U(3,\not C)$ color region. This feature leads to the {\bf
ans\"atz} that the photon field can be curved into the $U(n,\not
C)$-space with its energy increasing. So, we can add another
curving to the curvings of solely in $GL(n,\not C)$-space and/or
solely in $U(n,\not C)$-space. Now different sections of complex
spaces can be linked by the photon-field curving: $U(1,\not
C)\rightarrow U(2,\not C)\rightarrow U(3,\not C)$. In formalism,
this kind of curving can be interpreted by generalizing the
meaning of Eq. (7.3) in Ref. [1], with the indices $a, c$ being
not confined to one unitary group, but able to change continuously
among $U(1,\not C)$, $ U(2,\not C)$ and $ U(3,\not C)$.

Possibly, the dynamics of General Nonlocality also governs other
realms of nature, such as Thermodynamics or Hydrodynamics, if only
we find the appropriate space, as well as the metric forms and the
structure groups for the space. Then using geodetic line and the
field equation $R=0$ can mean all.

\end{document}